# Dynamic Solar Hosting Capacity Calculations in Microgrids

Z.S. HOSSEINI, A. KHODAEI
University of Denver
USA

E.A. PAASO, M.S. HOSSAN, D. LELIC
ComEd
USA

SUMMARY

Microgrids, as small-scale islandable power systems, are considered as viable promotors of renewable energy resources. In particular, microgrids can efficiently integrate solar photovoltaic (PV) as the deployment of this customer-deployed technology is growing in distribution grids. However, there is a limit on how much PV can be hosted by a microgrid. High penetration of PVs can potentially put the microgrid at operational risks including but not limited to over and under voltages, excessive line losses, overloading of transformers and feeders, and protection failure. To avoid such potential negative impacts, the concept of hosting capacity is introduced and used. The hosting capacity is defined as the total capacity of DERs that can be integrated into a given feeder/microgrid without exceeding operational restrictions and/or requiring system upgrades. Hosting capacity studies are primarily done based on steady-state analyses. However, in case of microgrids and when transitioning between grid-connected and islanded modes, dynamic operation becomes more restrictive than steady-state operation and thus is worthy of detailed investigation to provide a better understanding of the amount of DER that the microgrid can host. This paper examines this problem and provides extensive simulations on a practical test system to show its importance and merits.

Zohreh.Hosseini@du.edu

## 1. INTRODUCTION

A microgrid is a small-scale power system which consists of at least one distributed energy resource (DER) and one load and can perform in both grid-connected and islanded modes of operation [1]. The initial motivation to build microgrids was to streamline the integration of growing DERs in distribution grids. However, as the technology evolved, microgrids emerged as viable enablers of improved reliability, resilience, power quality, and energy efficiency, all offered through a sophisticated local intelligence and electricity supply capability [2], [3].

Microgrids are particularly of interest in integration of renewable energy resources, most commonly customer-deployed solar PVs. There is however a limit on how much PV a microgrid can accommodate, which is calculated as the lowest PV penetration level that causes violation of at least one technical limit [4], [5]. This penetration level is called "hosting capacity" in distribution grids, and the same concept can be used for microgrids. PV integration can have multiple impacts which include impact on thermal loading of distribution grid components, nodal voltage values, short circuit levels, and stability and power quality such as harmonics and voltage flicker [6]. There are also other impacts, such as possibility of reverse power flow due to high penetration and excess generation of PV. Another limiting factor is the microgrid islanding requirements which comes to the picture in the steady-state operation. To ensure microgrid's islanding capability, a large portion of the deployed DERs should be dispatchable, such as combined heat and power (CHP) or diesel generators. The dispatchable sources naturally reduce the amount of nondispatchable renewable resources to be deployed.

This paper performs a dynamic study to find hosting capacity of PVs installed in the Bronzeville Community Microgrid (BCM). The PV hosting capacity in this microgrid is achieved by analyzing the microgrid stability for different levels of PV penetration. In this process, the hosting capacity for the transition from the grid-connected mode to the islanded mode is analyzed. To investigate the impacts of diesel generator status, stability analyses are performed for both during the ON and OFF status of the diesel generator.

## 2. SYSTEM DEFINITION

The BCM, considered in this paper, is located in the Bronzeville neighborhood on the South side of Chicago, Illinois. The BCM deploys high-power PV which is controlled by the DOE-funded microgrid cluster controller and is connected to the DOE-funded 12 MW IIT Microgrid. A single-line diagram of Phase 1 of the project is shown in Figure 1 which presents the BCM's connection to the IIT microgrid and the ComEd's utility grid. For the sole purpose of the studies in this paper, it is assumed that PV and diesel generator are placed in bus 103.

Detailed BCM distribution model is implemented and analyzed using PSCAD [7] simulation tool to determine the maximum PV penetration level that can be hosted in BCM without causing instability or requiring additional investments. The model developed for BCM consists of 1) a voltage source for the main grid, 2) a diesel generator, 3) detailed solar PV model, and 4) loads and feeder impedances which are arranged based on a simplified topology of BCM. In addition, the model developed for PV system includes PV array, DC-DC Boost convertor, and voltage source converter.

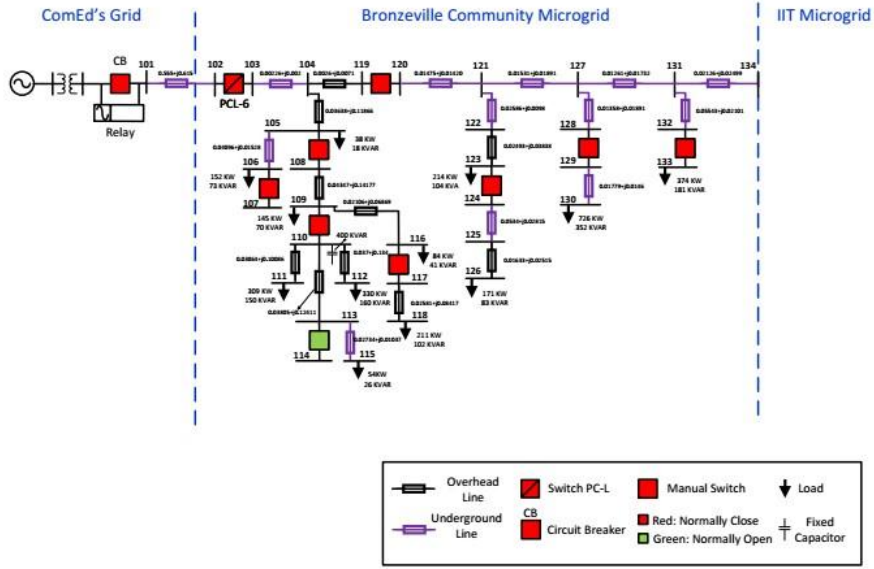

Figure 1. Single-line diagram of the BCM model.

The considered minimum and maximum daytime loads of the BCM are 1 MW and 2.8 MW, respectively, and the nominal voltage of the hosting feeder is 13.8 kV. Dynamic studies are performed for both minimum and maximum daytime loads to determine the hosting capacity of this microgrid to support the maximum penetration level of PV.

## 3. DYNAMIC STUDIES FOR HOSTING CAPACITY CALCULATION

A set of stability analyses are performed to find hosting capacity of the microgrid in case of transition from the grid-connected to the islanded mode and to investigate the impact of the diesel generator status on microgrid stability. In this respect, four different combinations of diesel generator status, i.e., ON or OFF, and high and low daytime load conditions are considered. The stability of the microgrid is analyzed when the microgrid is initially connected to the main grid and will be disconnected at t=3 s. In this paper, it is assumed that the excess generation by the diesel generator or PV is not curtailed and will be exported (sold back) to the main grid.

For each scenario, various levels of PV penetration from 0% to 60% of the load, with steps of 10%, are considered and the results for a pair of stable and unstable cases are plotted. To show the subsequent impacts of oscillations in the system, real and reactive power of the selected bus 104 are plotted since the aggregated power of three sources flows through this bus. The sources include the main grid, the diesel generator, and the PV unit.

**Scenario 1: Stability analysis for low load condition when diesel generator is OFF**

In this scenario, the BCM is in the grid-connected mode, initially. Next, it becomes disconnected from the main grid at t=3 s. Three cycles after the grid disconnection, the PV unit drops offline since it does not have a voltage source to synchronize with. The diesel generator starts up and comes online to supply the local load after six cycles (0.1 s) of the grid disconnection. After the diesel generator becomes online, the PV unit sets the diesel generator as its voltage source and synchronize with the diesel generator. The diesel generator experiences different levels of torque and speed oscillations depending on the microgrid load level and PV penetration level.

Figure 2 shows the real power, reactive power, speed, and torque of diesel generator, and real and reactive power of PV unit and bus 104, for low load condition when diesel generator is OFF, and the PV penetration is 10% of the load level, i.e., 0.1 MW. The diesel generator experiences speed and torque oscillations, however these oscillations are slightly damped. Damping oscillations of speed and torque of the diesel generator lead to damping oscillations on the PV unit. As the oscillations are damped, diesel generator and PV unit are able to pick up the load and microgrid remains stable.

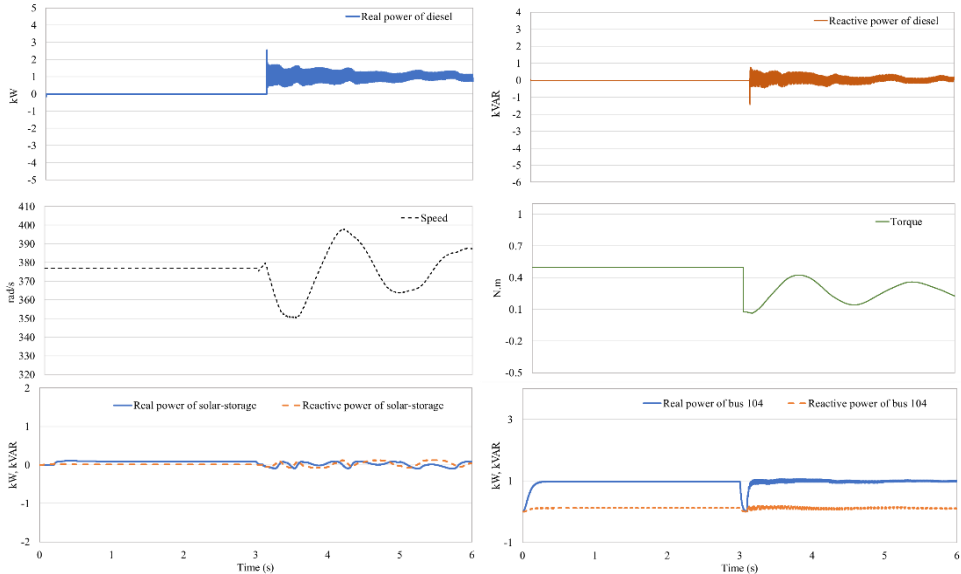

Figure 2. Real power, reactive power, speed, and torque of diesel generator, and real and reactive power of PV unit and bus 104, in scenario 1, when PV penetration level is 10% of the load level, i.e., 0.1 MW.

Figure 3 shows the real power, reactive power, speed, and torque of the diesel generator, and real and reactive power of PV unit and bus 104, for low load condition when diesel generator is OFF, and the PV penetration is 50% of the load level, i.e., 0.5 MW. As shown in this figure, diesel generator starts generating after getting disconnected from the grid while it experiences speed and torque oscillations which stay undamped. Subsequently, the PV unit which is tracking diesel generator as its voltage source, experiences different levels of oscillations. These oscillations may cause tripping in the system and lead to the instability.

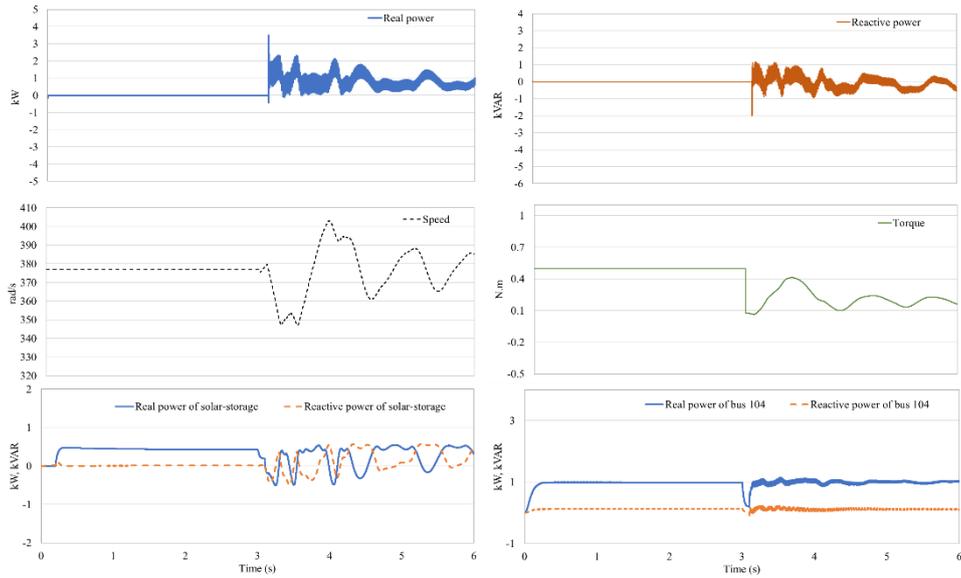

Figure 3. Real power, reactive power, speed, and torque of diesel generator, and real and reactive power of PV unit and bus 104, in scenario 1, when PV penetration level is 50% of the load level, i.e., 0.5 MW.

**Scenario 2: Stability analysis for low load condition when diesel generator is ON**

Like scenario 1, in this scenario BCM initially is connected to the main grid and then becomes disconnected from the grid at t=3 s. The diesel generator is initially online and synchronized. After getting disconnected from the grid, as the diesel generator is online, PV unit changes its voltage source from the grid to the diesel generator. In this scenario, the diesel generator experiences different torque and speed oscillations, depending on the microgrid load level and PV penetration level, which are much less than in scenario 1.

Figures 4 shows the results of scenario 2, when the PV penetration level is 0.1 MW. As shown in figure 4, the diesel generator initially generates 2 MW and PV unit produces 0.1 MW, and the excess power is exported to the main grid. After grid disconnection, different levels of speed and torque oscillations happen in the diesel generator, which cause damping oscillations in the system. During the test oscillations are damped, which means the microgrid remains stable after the transition.

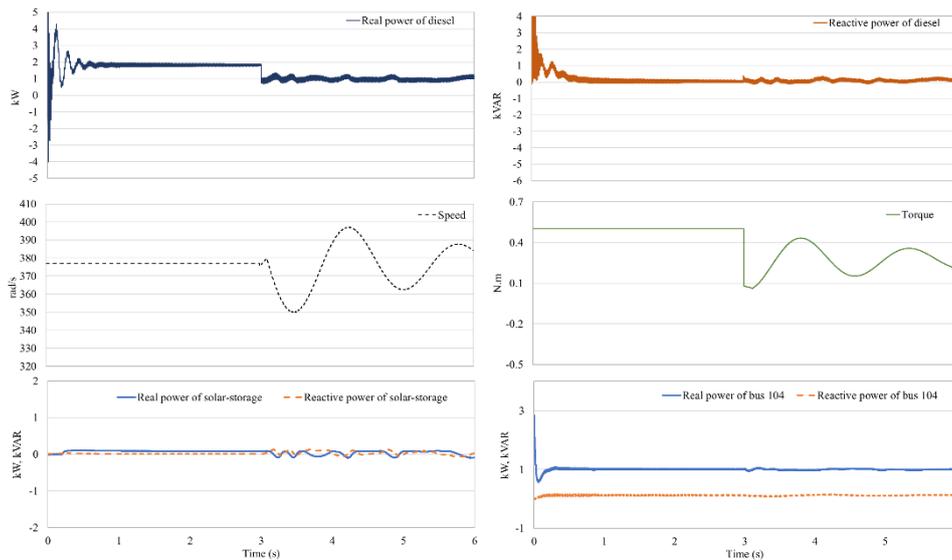

Figure 4. Real power, reactive power, speed, and torque of diesel generator, Real and reactive power of PV unit and bus 104, in scenario 2, when PV penetration level is 10% of the load level, i.e., 0.1 MW.

Figure 5 shows the results of scenario 2 when the PV penetration level is 0.6 MW. As shown in this figure, the speed and torque oscillations happening after t=3 s are undamped at the end of the test which make the system unstable.

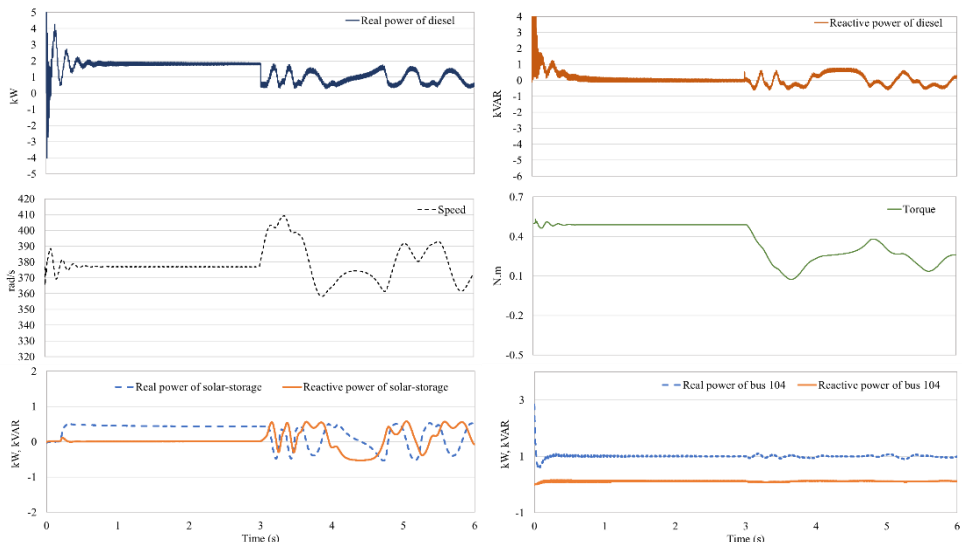

Figure 5. Real power, reactive power, speed, and torque of diesel generator, Real and reactive power of PV unit and bus 104, in scenario 2, when PV penetration level is 60% of the load level, i.e., 0.6 MW.

**Scenario 3: Stability analysis for high load condition when diesel generator is OFF**

This scenario assumes that load condition is high and diesel generator is initially OFF. The transition from the grid-connected mode to the islanded mode happens at t=3 s. Three cycles after the grid disconnection, the PV unit drops offline since it does not have a voltage source to synchronize with. The diesel generator starts up and comes online to supply the microgrid load after six cycles (0.1 s) of the grid disconnection. After the diesel generator becomes online, PV unit sets the diesel generator as its voltage source and synchronize with the diesel generator.

The diesel generator experiences significant levels of torque and speed oscillations depending on PV penetration level. In this scenario, based on the high load level which is 2.8 MW, different levels of PV penetration from 0% to 60% of the load level with steps of 10% are considered. As the load level is high and diesel generator is initially offline, after grid disconnection, diesel generator has to start up and significantly increase its generation. When PV penetration is low, diesel generator experiences huge oscillations and is hardly able to pick up the load. The following two examples demonstrate that in this scenario microgrid gets unstable in both low and high levels of PV penetration.

Figure 6 shows the real power and reactive power of diesel generator, PV unit, and bus 104 in scenario 3, when PV generation reaches the maximum power of 0.28 MW. After grid disconnection the diesel generator must go online and generate 2.5 MW to meet the load which causes significant oscillations. As shown in this figure, after t=3 s the diesel generator was able to pick up the load. The speed and torque oscillations happened to the diesel generator are undamped which means the microgrid is not able to recover and becomes unstable after the transition.

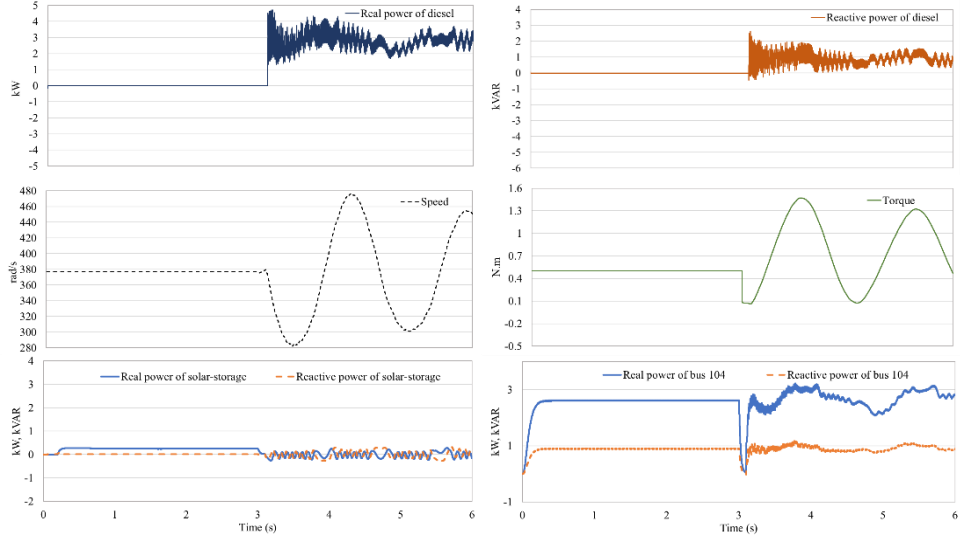

Figure 6. Real power, reactive power, speed, and torque of diesel generator, Real and reactive power of PV unit and bus 104, in scenario 3, when PV penetration level is 10% of the load level, i.e., 0.28 MW.

Figure 7 shows the results of scenario 3, when PV penetration is 0.85 MW. This figure illustrates that after grid disconnection, the diesel generator goes online to generate power. The load level is 2.8 MW and maximum power generation of PV is 0.85 MW which means the diesel generator must increase its generation from zero to approximately 2 MW. The diesel generator experiences severe increasing speed and torque oscillations. Due to these undamped oscillations, it is not able to pick up the load which causes tripping in the system and this the microgrid becomes unstable.

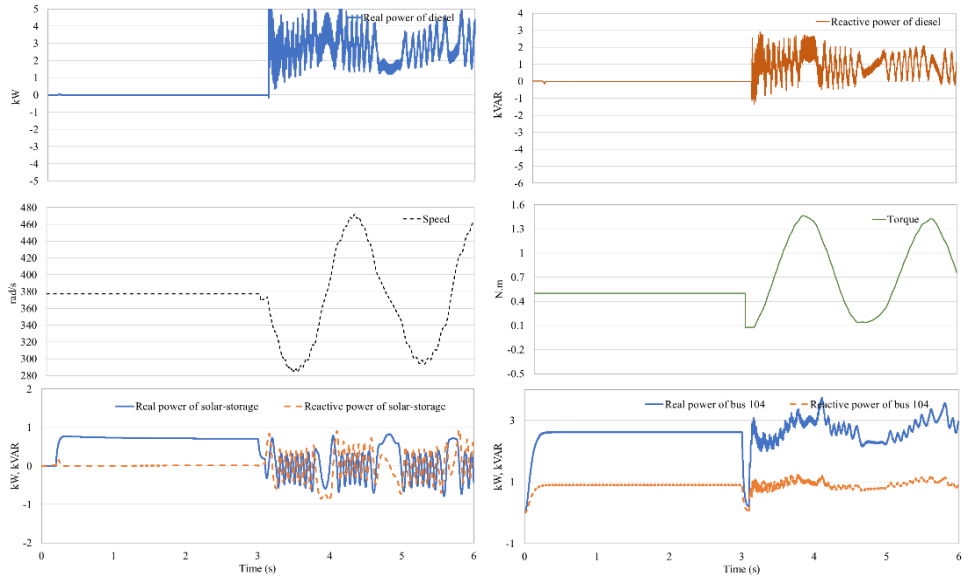

Figure 7. Real power, reactive power, speed, and torque of diesel generator, Real and reactive power of PV unit and bus 104, in scenario 3, when PV penetration level is 30% of the load level, i.e., 0.85 MW.

**Scenario 4: Stability analysis for high load condition when diesel generator is ON**

In this scenario, stability analysis is performed by considering high load condition while the diesel generator is initially online. After transition from the grid-connected mode to the islanded mode at t=3 s, the PV unit starts tracking the diesel generator as its voltage source. The diesel generator experiences different levels of torque and speed oscillations depending on the microgrid load level and PV penetration level. In this scenario based on the high load level which is 2.8 MW, different levels of PV penetration from 0% to 60% of the load level with a step of 10% are considered.

Figure 8 illustrates the results of scenario 4, where the PV unit reaches maximum power output of 0.85 MW. As shown in this figure, diesel generator is initially online and generates 2 MW, and the PV unit contributes to meeting the maximum daytime load of 2.8 MW by producing 0.85 MW power. After t=3 s, the microgrid becomes islanded and speed and torque of the diesel generator start oscillating. In this case, the speed and torque oscillations are significantly damped. At the end of the test, the diesel generator was able to regulate its frequency at 60 Hz. This figure also shows that the subsequent oscillations at PV output are also damped and microgrid is able to remain stable during and after the grid disconnection.

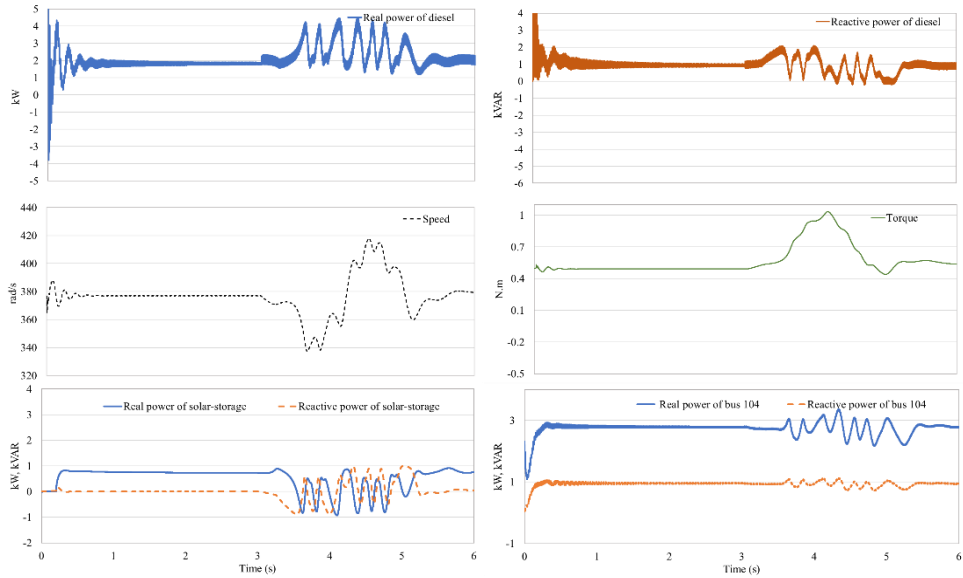

Figure 8. Real power, reactive power, speed, and torque of diesel generator, Real and reactive power of PV unit and bus 104, in scenario 4, when PV penetration level is 30% of the load level, i.e., 0.85 MW.

Figure 9 shows real power, reactive power, speed, and torque of the diesel generator, as well as real and reactive power of PV unit and bus 104, in the scenario 4 when PV penetration is at 1.4 MW level. As shown in this figure, after t=3 s, the diesel generator which was initially online, experiences different levels of speed and torque oscillations. Speed oscillation is damped at the end of the test and diesel generator regulates its frequency almost at 60 Hz, while the torque oscillation remains undamped. Note that the torque increases and it causes tripping and instability in the microgrid. These oscillations lead to the subsequent oscillations in the PV unit and other buses, as shown in Figure 9.

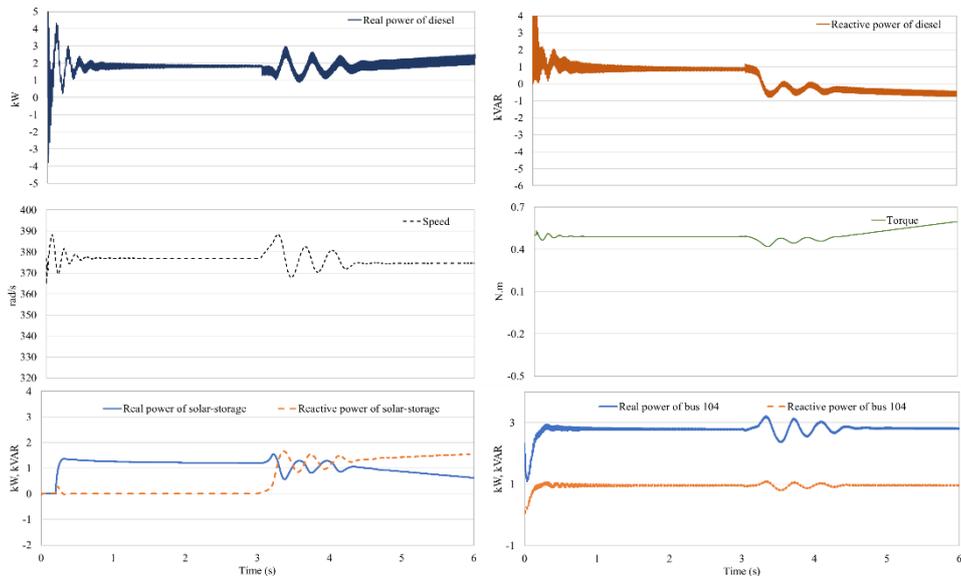

Figure 9. Real power, reactive power, speed, and torque of diesel generator, Real and reactive power of PV unit and bus 104, in scenario 4, when PV penetration level is 50% of the load level, i.e., 1.4 MW.

Table I shows the results of the stability analysis for four demonstrated scenarios for different levels of PV penetration. It should be noted that PV penetration levels are defined based on percentages of the load level in each scenario. For instance, PV penetration level of 10% of the low load (1 MW) is 0.1 MW, while PV penetration level of 10% of the high load (2.8 MW) is 0.28 MW. As shown in Table I, hosting capacities in low load scenarios are 40% and 50% of the load level, when the diesel generator is OFF and ON, respectively. The hosting capacity of high load scenario when the diesel generator is ON is 40% of the load level. However, when the diesel generator is OFF, the microgrid is unstable for all PV penetration levels which shows that it cannot host any PV under this condition. This table shows that by setting the diesel generator initially online, hosting capacity can be improved.

Table I. Results of stability analysis under different conditions.

|  |  | Low load (1 MW) | | High load (2.8 MW) | |
| --- | --- | --- | --- | --- | --- |
| Diesel status | | OFF | ON | OFF | ON |
| PV penetration level | 0% | Stable | Stable | Unstable | Stable |
| | 10% | Stable | Stable | Unstable | Stable |
| | 20% | Stable | Stable | Unstable | Stable |
| | 30% | Stable | Stable | Unstable | Stable |
| | 40% | Stable | Stable | Unstable | Stable |
| | 50% | Unstable | Stable | Unstable | Unstable |
| | 60% | Unstable | Unstable | Unstable | Unstable |
| | >60% | Unstable | Unstable | Unstable | Unstable |

## 4. DISCUSSIONS

Based on the cases studied, several points can be derived:
- It is significantly important to determine and keep adequate levels of spinning reserve in the microgrid dynamic operation as well as in supporting integration of higher levels of PV. The studied scenarios clearly show that when the diesel generator is ON the hosting capacity is increased, i.e., the microgrid has a grid-forming DER that can provide reserve if needed.
- If the microgrid is not equipped with adequate fast response grid-forming DERs, there is a chance of instability after the transition to the islanded mode that needs to be considered in both planning and operation of the microgrid.
- In this paper, the PV penetration level is changed with a 10% steps. Using smaller steps could result in a more accurate hosting capacity solution.
- The results obtained in this paper are for a predefined location of the solar PV units and it is assumed that these units are centralized in the mentioned location. By considering distributed integration strategy for the PV units and/or changing PV unit's location, the hosting capacity results may change.

## 5. CONCLUSIONS

In this paper, a dynamic study was performed to find the PV hosting capacity of a practical microgrid in case of transition from the grid-connected mode to the islanded mode. The impacts of initial status of the grid-forming DER on the stability of the microgrid was also considered and proved to be a key component in the studies. The stability analyses demonstrated that the microgrid hosting capacity, when the load level is low, can get to approximately 50% of the load for this specific system, while for high load scenarios the hosting capacity highly depends on the availability of the grid-forming DER. In other words, when the grid-forming DER is initially online, the microgrid is more stable after being disconnected from the grid and the hosting capacity is improved, thus highlighting the fact that microgrid dynamic hosting capacity is exceedingly dependent on its operational conditions.